# Improving transferability between different engineering stages in the development of automated material flow modules

D. Regulin, T. Aicher and B. Vogel-Heuser

*Abstract*—To improve the flexibility and robustness of the engineering of automated production systems (aPS) in the case of extending, reducing or modifying parts, several approaches propose an encapsulation and clustering of related functions, e.g. from electrical, mechanical or software engineering, based on a modular architecture. Considering the development of these modules, there are different stages, e.g. module planning or functional engineering, that have to be completed.
A reference model that addresses the different stages for the engineering of aPS is proposed by the Automation Markup Language (AutomationML). Due to these different stages and the integration of several engineering disciplines, e.g. mechanical, electrical/electronic or software engineering, information not limited to one discipline are stored redundantly, increasing the effort to transfer information and the risk of inconsistency. Although data formats for the storage and exchange of plant engineering information exist, e.g. AutomationML, fixed domain specific structures and relations of the information, e.g. for automated material flow systems (aMFS), are missing. This paper presents the integration of a meta model into the development of modules for aMFS to improve the transferability and consistency of information between the different engineering stages and the increasing level of detail from coarse-grained plant planning to fine-grained functional engineering.

*Note to Practitioners*—The engineering of automated production systems (aPS) is a highly interdisciplinary procedure that requires the interaction of all participating people. However, communication during the engineering process is based on the exchange of information, orally or by documents. Additionally, many influences during the different design stages, i.e. time periods, cause the change of parts or parameters. Hence, the neighboring engineering disciplines that employ these data need to be updated regarding the changes. Consequently, one challenge during the engineering of aPS is the data exchange as well as the guarantee for a consistent and complete description of the system. This contribution presents a methodology for a data exchange between the different tools applied during the engineering cycle in the design process of aPS. Therefore, the reference process proposed by the Automation Markup Language (AutomationML) as well as the associated data model for data exchange are applied. In contrast to the current state of the art, our approach additionally proposes the integration of a meta model of the aPS in order to identify all engineering data required for a complete description of an aPS. Furthermore, the consistency is guaranteed by references between the meta model elements, which organize and provide the required information for the engineer at a certain stage through links to the needed parameters or documents of neighboring disciplines. In this context, sub-elements of the meta model provide interfaces to the specific engineering tools. As a result, the effort for information exchange decreases and a higher level of data consistency could be reached. In future research, we will examine measurements to quantify the increased efficiency compared to further current engineering approaches.

## I. INTRODUCTION

Developing complex automated production systems (aPS) and distribution centers by integrating different engineering disciplines, e.g. mechanical, electrical/electronic or software engineering, is a rising challenge for plant manufacturing. In addition to the requirements of high dependability and throughput, aPS – living up to three decades of operation – have to consist of a flexible adaptable structure based on the various incremental versions as a consequence of sequential evolution over time [1].

Dealing with the requirement to extend, reduce or modify parts of the aPS dynamically, an established approach of planning and constructing whole aPS or automated material flow systems (aMFS) describes their decomposition in small subsystems, i.e. modules [2]. In this approach, to reduce the effort for the software-engineering, predefined software components for these modules exist and can be interconnected to each other. Therefore, the complexity of the development can be broken down to the development of subsystems (modules) by encapsulating their description and their related information, and, subsequently, the composition of multiple modules for the design of an aPS, e.g. aMFS. Therefore, generic interfaces, which enable the control engineering of the module, but also the application of higher level functions, e.g. the material flow of a whole logistic center, are involved.

Considering the development of these modules or aPS itself, different engineering processes are required to complete an aPS. Consequently, the whole development process of the aPS is divided into stages, i.e. time periods. Based on the requirements of the engineering process, e.g. the domain or complexity of the system, as well as the working methods of the involved developers, e.g. interdisciplinary vs. separated engineering approaches, the particularly most suitable engineering process is chosen. A reference model for the engineering process of aPS is proposed by AutomationML (AML) [3],[4]. However, considering engineering processes for mechatronic systems, e.g. aMFS, a generic separation into the concept design, e.g. factory or module planning consisting of electrical, process, logistic or quality assurances, and detail planning, e.g. functional engineering consisting of mechanical, electrical and software parts, is required [5]. Hence, a system and its modules are developed separately in different engineering domains, resulting in an increased effort regarding redundant work, complexity and error-proneness resulting from the missing transferability of the information and description documents. To handle the complexity and, thereby, reduce the effort regarding development in functional engineering, in this paper a meta model to support the development procedure for the functional engineering and especially software-engineering for automated material flow modules (aMFM) is proposed. Thus, based on the results of the plant planning and the dependencies of the different engineering disciplines, the relations of information are organized according to the AML reference model for the engineering process, on the one hand, and in reference to the knowledge of the system architecture provided by a meta model, on the other hand. Consequently, a fixed structure of the information and the references between them stored in a consistent data format, e.g. AML, can be derived from the meta model.

This paper is organized as follows: In the following section, the requirements addressed by our introduced procedure model of functional engineering are specified. Subsequently, in section III, state of the art technologies and approaches are evaluated against these requirements. Our approach is introduced in section IV. Based on that, we show our solution for a procedure model of the functional engineering in section V. This paper concludes with a summary and a discussion of the results and future work in section VI.

## II. REQUIREMENTS

To reduce error-proneness and effort in the interdisciplinary functional engineering process of aMFS, several requirements regarding the support exist and have to be fulfilled by a model driven approach.

*Aggregation of module-specific documents of the different planning stages (R1)*

During the engineering process, documents related to a module are created. According to the AML-process model, these documents can be divided into planning stages, i.e. plant planning, functional engineering and commissioning/production. In addition, different domains, e.g. mechanical, electrical and control engineering, are involved and have specific viewpoints on the system [6]. Considering these different stages and domains, a multitude of information related to these disciplines exist. Examples are sensors or actuators, which are first planned roughly in the plant planning stage. However, the generated information is used for further design in mechanical, electrical and software engineering. Consequently, multiple engineering disciplines and stages are dependent on the information of these components. Hence, the communication and exchange of information about the module are prerequisites for an efficient engineering process. Thus, an aggregation of the created documents of the different disciplines is required.

*References among the module-specific engineering information and assignment to the related discipline (R2)*

Along with the creation of documents, engineers rely on information from different disciplines. For example, the control engineer develops the aMFS's software based on information about sensors, actuators as well as the position and electrical connection to the programmable logic controller (PLC), which were planned by the mechanical and electrical engineers [7]. Hence, inside the aggregated information on a module (cf. R1), assignments between the parameters and properties inside the different disciplines have to be integrated. Therefore, the information of the single documents have to be referenced by each other, e.g. position, signal, connector and software function of a sensor, between different stages and disciplines.

*Meta model for the assignment of the document's information to the correct module part (R3)*

Based on the aggregation and references among the documents, engineers are able to acquire the information required for the specific development task, e.g. a part or piece of software. To minimize the search time for specific information, the documents as well as the references among them have to be structured. Since the architecture of the aPS depends on the specific domain, e.g. aMFS, the structure and references of the documents have to follow the particular architecture of these domain's modules. Meta models of such aMFS provide the knowledge of the architecture and existing references among the module's parts and software. Consequently, a structured organization of engineering data can be achieved through assignment to a given meta model and inheritance of its references. Hence, the information of the different documents is assigned to the related classes in the meta model and to the module's elements, e.g. parts or software.

*Procedure for a structural complementation of the information required for the functional engineering (R4)*

Beside the architecture, the meta model provides the set of parameters and properties, which are required for the complete description of a plant module. This information is the main result of the different engineering disciplines and is a subset of the content of the created documents. Since the semantics and syntactics of the parameters can be different in the documents related to the disciplines, the meta model suggests rules for semantics, syntactics and the required data. Consequently, the standardized definition should enable the data acquisition of various disciplines as well as different stages of engineering. Consequently, all information required for the functional engineering should be covered by the meta model.

## III. STATE OF THE ART

In this section, model-driven engineering approaches and procedures, which are common methodologies for developing automated systems, e.g. aMFS, are analysed regarding the introduced requirements.

Fan et al. show how geometry-related data can be exchanged between models, especially the automatic generation of computer-aided design (CAD) models [8]. Additionally, based on model comparison even applied to various platforms, conflicts can be avoided. However, the methodology is focused on the structural data of the system and the author emphasizes the need for the integration of other engineering disciplines, e.g. functional behaviour (cf. R4).

Moscato et al. present a Solution Process Definition Language, which can be applied to generate executable processes for heterogeneous systems automatically [9]. The author provides an evaluation based on a hospital infrastructure, which also includes the perspective of exchanged information. However, the solution model does not consider engineering processes or a reference architecture for aMFM (cf. R4).

Liqing et al. presents a design implementation of fixture designs (FD) and analysis systems based on a service oriented architecture to enable the collaboration of designers across the globe [10]. The associated information models are based on the extensible markup language (XML) standard and support the information exchange between the disciplines FD and analysis. Since the approach only focuses on parts of the engineering process, i.e. the mechanical engineering, the scope has to be extended about the neighboring disciplines (cf. R1).

The relation between the demand for final products and the operations is represented by a model from Cunha et al. [11]. Based on a mathematical model, the lowest cost consumption of the different operations in a supply-chain is determined. Consequently, the information of the different stages are exchanged. Nevertheless, the model does not support the references between engineering stages and the associated documents (cf. R2).

Boschian et al. present an integrated system (IS), which can be applied to manage Intermodal Transportation Networks at operational and tactical levels [12]. The knowledge base of the IS is defined by a reference model that foresees the system behaviour and provides the data for management strategies. A meta model defines the structure of the systems and processes based on the architectural knowledge. The evaluation has shown the advantages of the metamodeling approach regarding application, updating and changing. Our approach considers the results of Boschian et al. and transfers and extends the methodology into the field of systems engineering for aPS (cf. R1).

How complex mechatronic systems can be modeled in an object oriented way using the Unified Modeling Language (UML) is presented by Secchi et al [13], [14], [15]. The model considers the system components as well as the associated control software and interaction by exchange of information. Hence, the approach enables the system description regarding structure and architecture, but does not enable the exchange of information within the different engineering stages based on a process reference model (cf. R3, R4).

A consistent engineering information model for mechatronic components in aPS is presented by Aicher et al. [24]. Therein, AML is used to improve the transferability and consistency of information from the manufacturer of components to the software engineering of aPS (cf. R1, R2). However, an integration of a meta model for assignment of the information to different domains or stages as well as the integration of a procedure model are not proposed (cf. R3, R4).

Tab.1: Review and classification of existing research approaches for model based engineering approaches in the field of aPS

| | Aggregation of module specific documents (R1) | References among information (R2) | Meta model for correct assignment (R3) | Structural complementation of for the functional engineering (R4) | Domain | Disciplines | Modeling language |
|---|---|---|---|---|---|---|---|
| Fan et al. [8] | + | - | + | - | Multi-domain | Mech. engineering | SysML |
| Moscato et al. [9] | + | - | + | - | Formal modeling/ healthcare | Workflow-planning | Solution process definition modeling language (SPDL) |
| Liqing et al. [10] | + | + | - | - | - | Mech. engineering | CAD, XML |
| Cunha et al. [11] | - | + | - | - | Supply-Chain, automotive | Assembly, planning | Mathematical model |
| Boschian et al. [12] | + | - | + | + | Transportation networks | Transportation management | UML |
| Secchi et al [13], [14], [15] | - | + | - | - | Industrial automation | Requirements and systems engineering, software architecture | UML/ SysML |
| Anacker et al. [16] | - | + | + | + | Transportation | Requirements and systems engineering | - |
| Black et al. [17] and Doukas et al. [18] | - | - | - | + | Material flow systems | Softw. engineering | IEC 61499 |
| Biffl et al. [19] and Schleipen et al. [20] | + | + | - | + | Industrial automation | Mech., electr., softw. engineering | AutomationML |
| Estevez et al. [21] | + | - | - | + | Industrial automation | Systems and softw. engineering | AutomationML, PLCopen |
| Wehrmeister et al. [22] | - | - | - | + | Automation systems | Softw. engineering | UML |
| Schröck et al. [23] | + | - | - | + | Automation and embedded systems | Requirements, mech., electr., softw. engineering | Software Platform Embedded Systems 2020 (SPES 2020) |

Wilfried et al. emphasised the demand of reconfiguration in logistic chains, due to the adaption to changing market demands and, therefore, presented a configuration model for a logistic chain [25]. However, since aMFM are a subset of the chain, the model-based design has to enable the adaptation of modules concerning the related changes by an appropriate meta model not published by Wilfried et al. (cf. R1, R4).

A model-based design approach for mechatronic systems, is presented by Anacker et al. [16]. The design process uses partial models organized in a "Reconfiguration Structure Matrix" and an "Aggregation-DSM". However, a structure for a specific system as well as functional aspects are not proposed (cf. R3).

Black et al. [17] and Doukas et al. [18] present an approach for model-driven engineering based on the IEC 61499 function block paradigm. However, a general meta model or procedure model for aMFM is not provided (cf. R3).

Biffl et al. present an approach for the information exchange of different engineering tools in order to improve the cooperation between the different disciplines based on the AML data model [19]. Additionally, Schleipen et al. realised an approach for "plug&produce" based on OPC-UA and AutomationML [20]. However, a meta model for the description of a module and organization of parameters and references, especially for aMFM, is not proposed (cf. R4).

Estevèz et al. [21] present an engineering approach applying AutomationML as an exchange format during the engineering cycle. Subsequently, the transformation into PLCopen-XML enables the code generation and even the creation of reconfigurable automation systems [29], [30]. Nevertheless, a meta model for a complete plant description as well as the references to related documents are not presented (cf. R3, R2).

Wehrmeister et al. present an approach for the UML-based design of real-time and embedded systems for automation applications [22]. The proposed design flow covers activities from earlier phases up to system implementation using a predefined target platform. However, the approach does not provide the aggregation of documents or an appropriate meta model for aMFM (R1, R3).

Schröck et al. present a meta model that considers relations between parts of a plant. Based on these links and several separation rules, a general engineering model for automation system is introduced [23]. Additionally, the authors emphasise the demand for considering the discipline-specific models, which is a main point of our approach. Nevertheless, a specific structure for a plant or a module is not presented.

The demand for syntactical and semantic correctness of variant-rich aPS, is presented by Vogel-Heuser et al. [26]. They provide an interdisciplinary survey on the challenges and state of the art in the evolution of aPS and identify that the challenge of the evolution of aPS is the coverage of all possible solution set variants by a description or tool. Further, Maga et. al [27] recognise that engineering tools in the aPS has its own significant benefits, but are insufficient for domain engineering in general and for reuse, in particular. Hence, currently available engineering tools cannot consistently cover the dependencies between software and hardware components.

Additionally, Feldmann et. al. [28] provide results that nowadays module structures in different disciplines of aPS, i.e. software engineering and electrical engineering, significantly differ. A consistent meta model to aggregate the documents of different planning stages and disciplines is lacking (cf. R3).

Subsequently, there are several approaches for the description of mechatronic systems during the engineering stages and even for saving the information according to a common data model, e.g. AML. Nevertheless, a model-based methodology for exchanging information according to a reference engineering process for aMFS, which considers the module architecture to organize the complete module description, is not available yet.

## IV. Concept

This section describes how the model-based description can be used to exchange information provided by engineering documents between the engineering disciplines and stages for the development of the functional description. Therefore, the reference model for the engineering process of aMFM proposed by AML is used. Before the concept is illustrated, a brief overview of the meta model AutoMFM [31] which is based on the material flow description by Wilke [32] is given.

To improve the applicability of the AutoMFM and describe the different modelling aspects of aMFM explicitly, a separation of the module information into the (sub-)classes *general description*, *status description*, *function description*, *module interface description* and *control description* is provided. Since the level of detail of the aMFM can be modelled in every (sub-)class separately, different engineering steps, which exist during the development process describing the system in different granularity, are supported. The meta model mainly consists of technical elements, which are assigned to a model's structural and functional description. Each class is divided into further (sub-)classes that represent a detailed view on the module, e.g. sensors, actuators or dimensions (see Fig. 1). Subsequently, the engineer can add module-specific parameters if a new module document is instanced. Since the required model data is included in the different engineering documents, the engineer additionally adds the reference to the related document. Consequently, different information is aggregated by the meta model for aMFM (R1) [31].

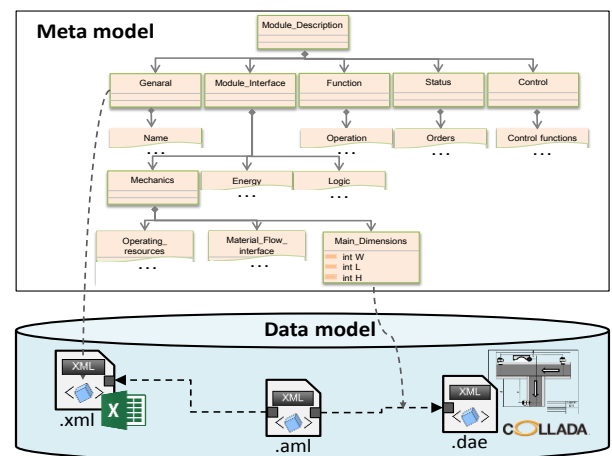

Fig. 1: References between different information derived from the introduced meta model AutoMFM [31]

In order to describe the basic information of the aMFM, e.g. the dimensions or type of module, the (sub-)class *general description* is contained in AutoMFM. Variable information, e.g. the current occupancy rate, is not contained in this (sub-)class. Instead, a general overview based on the static information of the module, which is particularly essential during engineering and planning, is given, e.g. name and identifier.

In addition to static information from the *general description*, runtime values of variable attributes, which is essential information for different disciplines, e.g. software engineering, are contained in the (sub-)class *status description*. Thus, a current time- and system-dependent representation of the aMFM, e.g. containing information of the operating mode or energy consumption, can be given. To describe the coherence of the different parameters, mathematical equations or logical descriptions can be modelled in further classes, e.g. *control description* or *function description*.

To describe the software, but also the hardware, for controlling the aMFM, the (sub-)class *control* is contained in AutoMFM. Therein, logical functions of the module specified for the programming languages of the IEC61131-3, but also the corresponding variables to store local function dependent information, can be modelled. Further, a mapping list of inputs and outputs to the corresponding variables to read sensor values and to control actuators is allocated. Considering the deployment of the software on different field devices, e.g. PLC, additional information from the device described in the platform information, e.g. type of bus coupler, can be modelled in this (sub-)class.

To represent logistic functionalities, e.g. conveying or buffering, the (sub-)class *function* is contained in AutoMFM. Hence, the material flow of the TU, e.g. the route of the TU or the priority at crosses, can be designed. Further, the interactions between neighboring aMFM or the selection of specific tasks, segmented into the three categories of material flow, handling and waiting, can be described.

To interact with neighboring aMFM, e.g. to hand over TU, interfaces for transferring information between different, e.g. neighboring, modules are necessary to model aMFS. Hence, a further (sub-)class named module *interface description* is contained in AutoMFM. To improve the consistency of the different kinds of interfaces, a standardisation of the interfaces is realised through the structure of this (sub-)class. Further, a common working space, called an interaction space, is modelled in this (sub-)class in order to aid in the interaction of different modules with one another.

After introducing AutoMFM, as a prerequisite for our concept, different information regarding an aMFM, e.g. dimensions, logic or structure, has to be stored in a data format consistently. Therefore, AutomationML (AML), which is a neutral data format based on XML, is chosen. However, developers using AML are exempted from conditions involving the structure or references of the module-based information. Thus, several approaches for storing information about one module in AML are feasible, but a consistency specification based on a meta model is lacking (see Fig. 1)

Considering the AML process model, only the sequence of the different engineering stages is provided; but the information flow between the plant planning and functional engineering is missing. Hence, the introduced meta model autoMFM [31] for aMFM is added in functional engineering to store all necessary information from the plant planning stage consistently (dashed lines in Fig. 2).

*A. Material flow specification*

The basic specification of the module's behavior is defined by the planning documents of the material flow process through the plant planning states *Process Planning, Logistics Planning and Electrical Planning* (Fig. 2). Since a modular logistic plant consists of several single modules, the process and logistic planning results contain coarse grained information, which is related to routing and the functionality of single modules (AutoMFM class *general description*). Additionally, the relation for the control of actuators dependent on sensor values and higher level information are specified (AutoMFM class *control*). Hence, this information is related to the description for the logistic functions as well as the interface, which are both sub elements of the material flow element (AutoMFM classes *function* and *interface description*). Based on the references between the elements of the meta model, the information about the dependencies among sensors and actuators as well as the process is also available in the control elements.

*B. Mechanical engineering and Layout/ Process refining*

Based on the material flow specification for logistics, process, component and coarse-grained electrical planning, the mechanical description of the logistic plant can be designed using an appropriate CAD tool. Since we assume a modular plant architecture, the plant engineer composes the plant by assembling the correct module instances. Along with the material flow functionality, the module engineer, on the other hand, specifies the detailed mechanical properties of each single module, e.g. the position of sensors, actuators and mechanical module interfaces. Since mechanical engineering is the prerequisite for the manufacturing and assembling of the module's components, the results of this engineering stage are CAD and computer-aided manufacturing (CAM) files for generating numerical control programs for machines, production drawings and fine-grained layouts (AutoMFM classes *general, functional, interface*).

The presented approach in this paper extends the possibilities for usage of the created documents for the functional engineering stage of a module. Therefore, the created documents are assigned to the sub elements of the aMFM meta model, e.g. fine-grained layouts or bill of materials (R3). The meta model specifies the required parameters for each sub element, e.g. a sensor, which have to be defined in order to describe the component, e.g. position and type of sensors and actuators as well as dimensions (R4). Additionally, references to the sub-elements of control engineering, e.g. inputs and outputs, as well as to the sub-elements of the electrical engineering, e.g. number and types of sensors, are used to publish the information in the required model elements (R2).

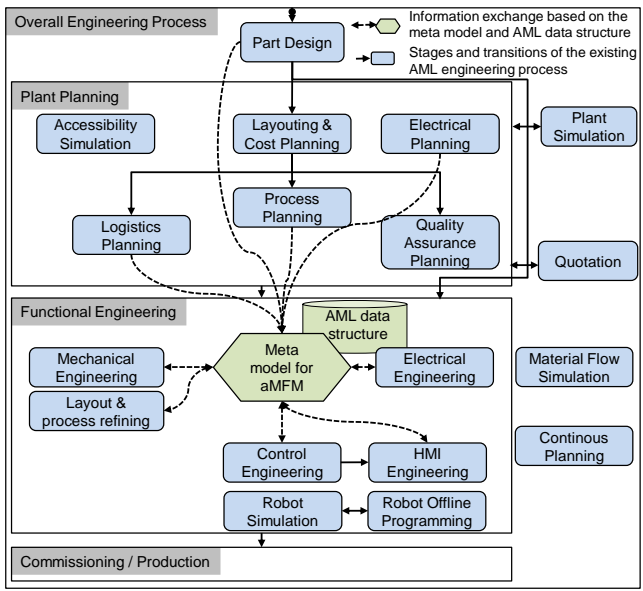

Fig. 2: Information flow of the AML- engineering states into the meta model ([33], [5])

*C. Electrical engineering*

Based on the coarse-grained planning documents of a plant's planning stage, the detailed electrical models are designed within the functional planning stage. Analogous to the mechanical design, the fine-grained electrical specification is supported by computer aided design tools for electrical planning (ECAD). Within this stage of engineering the inputs and outputs of the PLC and their connection to sensors and actuators are specified (AutoMFM classes *general, control*). Additionally, communication and safety facilities are planned (AutoMFM class *control*). Hence, the mechanical parameters, e.g. lists of sensors, actuators and their descriptions, are required to determine the connections to each other. This information is available through the linked parameters in the sub-elements of the meta model (R4). Further information, e.g. layout or technical drawings are provided through references to the assigned documents (R3).

Along with the usage of the provided data of the meta model, the electrical planning generates documents and information related to the required parameters in the meta model, which is a prerequisite for later engineering states, e.g. input and output addresses of sensors and actuators for control planning (AutoMFM class *control*).

*D. Control and Human Machine Interface (HMI)-engineering*

The control and HMI engineering specify the behaviour of a module in a formalized way to subsequently generate executable control code for PLC or robot controllers, e.g. by state machines. Since the material flow planning is part of the plant planning stage, constraints for the detailed design of module functions are given. Hence, for the development of the software functions, information of other engineering disciplines is required (R1), e.g. the distances between sensors and actuators as well as their position, to define the correct sequences for the system's states and failure recognition based on the time-dependent behavior (AutoMFM classes *control, status*). Since the control sequence of actuators is also defined within the system's states, the functional description has to consider constraints, e.g. the movement of an actuator. Thus, the application engineer needs information about the system's layout as well as specific parts assigned to the (sub-) classes of the module model (AutoMFM classes *general, interface, function*) (R3). Analogous to the mechanical information, electrical parameters are required to trigger actuator outputs and read sensor values within the software functions (R4). The description has to consider the mapping between electrical ports and logical addresses as well as the data types of signal values (AutoMFM class *control*). The HMI is closely related to the control software engineering and requires graphical layout information and a representation of the dynamic status information of sensors and actuators. This information and these mapping tables can be generated automatically based on the references of the module model elements (AutoMFM classes *status, control*) (R2).

*E. Rules for semantics, syntactics and the required data*

To deal with the procedure for a structural complementation of the information, rules for semantics, syntactics and the required data are necessary (R4). In addition, to improve the clarity and applicability of the rules by the engineers, not proprietary rules, but standardized rules have been considered instead. Hence, our approach uses AML as a data format and applies the AML standard role-class *AutomationMLBaseRoleClassLib* and standard interface class *AutomationMLInterfaceClassLib* to describe the abstract functionality without defining the underlying technical implementation and the relations between the different (sub-)classes in AutoMFM, respectively (see Fig. 3). For a more detailed description of the functionalities and interfaces, AML provides further standard role classes, e.g. *AutomationMLCSRoleClassLib*, or interface classes, e.g. *CommunicationInterfaceClassLib*, which are derived from the standard role and interface classes and can be used for storing the data of the modelled aMFM. Based on the mapping of the AutomationML *SystemUnitClassLib*, which are (pre-)defined in AutoMFM, objects instantiated in different domains can be mapped to each other. Additionally, using the corresponded roles enable the direct mapping of the objects (see Fig. 3).

The following formulas show an excerpt from the formal rule-definition for the semantic and syntactic description of the information modelled in AutoMFM. The introduced functions $I_1$, $R_1$, $I_2$, $R_2$ $I_3$ and $R_3$ map the information allocated in the meta model to its feasible AML role or interface description.

| Meta Model AutoMFM | | Roles and interfaces in the data format AutomationML |
|---|---|---|
| $I_1$: {Control.ContolFunction} | → | {AutomationMLBaseInterface.AttachmentInterface, AutomationMLBaseInterface.ExternalDataConnector.PLCopenXMLInterface} |
| $R_1$: {Control.ContolFunction} | → | {AutomationMLCSRoleClassLib.ControlEquipment} |
| $I_2$: {Function.LogisticFunction} | → | {AutomationMLBaseInterface.AttachmentInterface, AutomationMLBaseInterface.ExternalDataConnector.COLLADAInterface} |
| $R_2$: {Function.LogisticFunction} | → | {AutomationMLExtendedRoleClassLib} |
| $I_3$: {General.Identification} | → | {AutomationMLInterfaceClassLib.AutomationMLBaseInterface, CommunicationInterfaceClassLib} |
| $R_3$: {General.Identification} | → | {AutomationMLDMIRoleClassLib.DiscManufacturingEquipment, AutomationMLExtendedRoleClassLib} |

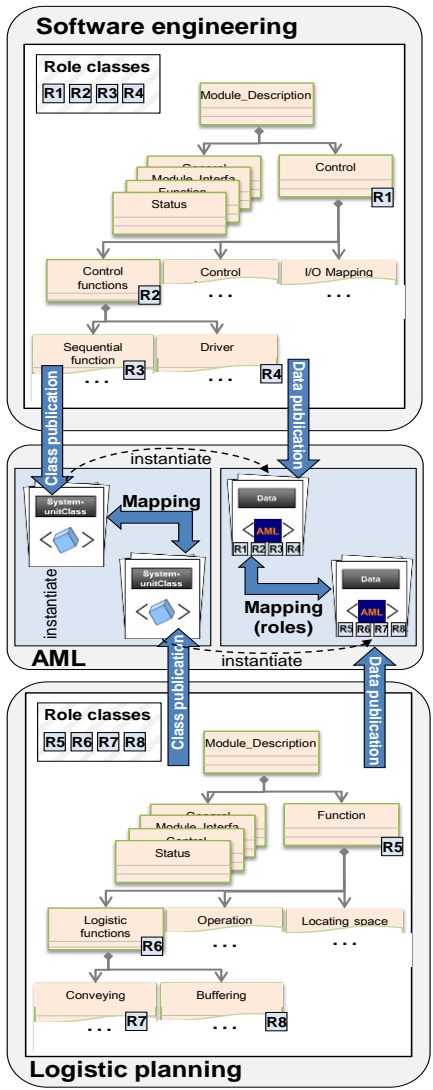

Fig. 3: Excerpt for semantic referencing of documents between different disciplines

Subsequently, based on the AML standard and the appropriate assignment of the AutoMFM elements, a formal description for the exchanged engineering information can be achieved.

## V. APPLICATION EXAMPLE AND EVALUATION

The following section presents the application of the model-driven development process based on the proposed meta model architecture for a simple exemplary logistic module, i.e. a T-junction.

The logistic T-junction module is employed to convey transport units (TU) and to perform dedicated TU-specific changes of the route. Therefore, the module consists of a main conveyor belt, which can convey TU in one direction. Additionally, one light barrier detects TU at the input of the conveyor belt and one, which detects TU at the output *"output_1"*. The second conveyor belt also consists of an input as well as an output *"output_2"* and can also feed TU in one direction. To detect TU at the output, there is also an additional light barrier installed at the second conveyor. To route TU from the main conveyor belt to *output_2* of the second conveyor belt, a pneumatic actuated switch is used. The position of the switch is monitored by two sensors. The material flow description specifies the behavior of the module dependent on the transport requests and sensor values on a coarse-grained abstraction level (Fig. 4).

If a TU enters the conveyor (1.0) and an ordering request for output 1 exists (1.1), the conveyor 1 has to be activated (1.2) until the TU has reached *output_1* (1.3). In case there is an order request for *output_2* (2.1) both the conveyors and the switch have to be activated (2.2). When the TU has reached the output of conveyor 2, both conveyors and the switch have to be deactivated (2.3). Subsequently, a reference of the document is stored in the class "logistic function" of the module model (R3). The logistic functionality and behavior description of the T-junction was described in a Pert-Chart diagram. Afterwards, to connect the information to AML, the Pert-Chart diagram was converted into an intermediate modelling language (IML) description and, subsequently, into a PLCopen XML format (see Fig. 5).

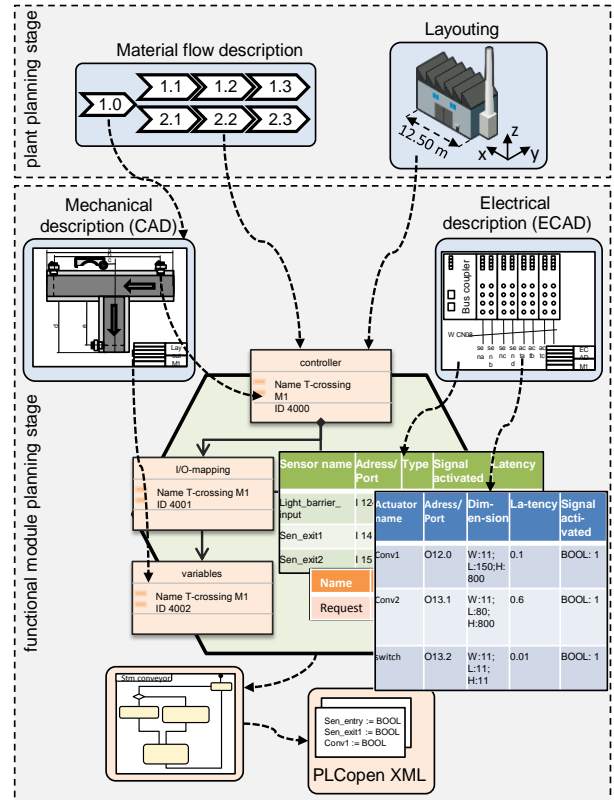

Fig. 4: Application example of the introduced functional engineering workflow

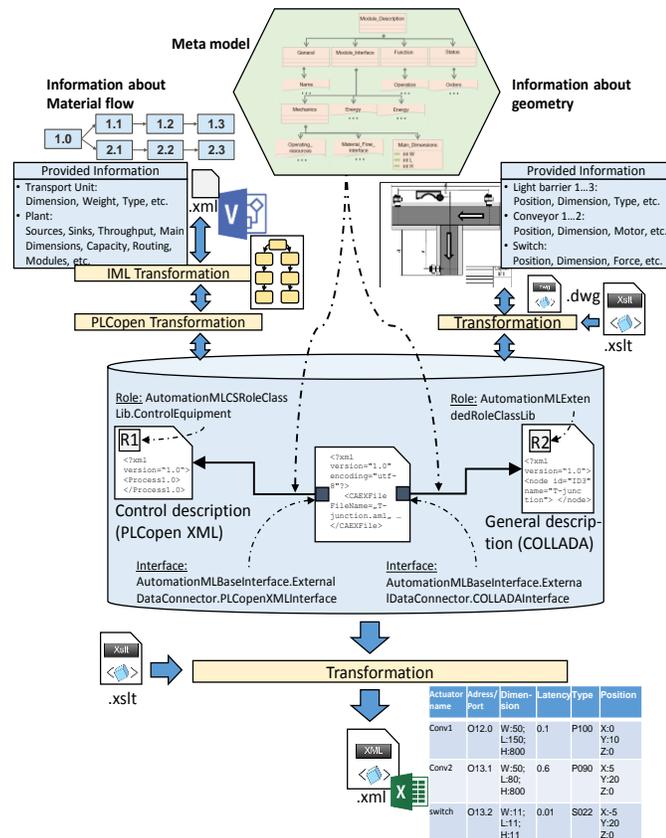

Fig. 5: Combining planning and geometry information based on the introduced meta model AutoMFM

In addition to the behavior description, the detailed layout data is provided by the CAD model, e.g. by Collada. Since the Collada-CAD tool supports XML-based conversion according to the specification in the ".xslt-file" of the geometric files (.dwg) into the AML format, the information can be stored in the AML data structure (see Fig. 5). The meta model in our approach defines the references between the elements, which require the information of another discipline, e.g. the main dimensions in the layouting, mechanical or control engineering, in order to describe the module completely (see Fig. 5) (R4).

One way to complete the information required for a module description that cannot yet be collected automatically through interfaces to special tools from the engineering disciplines, is the request of missing parameters of the associated engineer by tables. Therefore, the model data can be converted by XML-operations (.xslt-file) into a common tool for tabular calculations, i.e. Microsoft Excel. In addition to the parameter values, the related documents can be referenced by their name and path on a server used for data exchange. Therefore, the information is stored by the data type string in the AML data model.

The data of the other disciplines are converted and stored through the interfaces of the other model classes in an analogous way, e.g. the excerpt of the sub element conveyor represented by the information: name: Conv1; type: P100; main dimensions: (50,150,800) mm; latency: 0.1 sec; position: (0,10,0) (see Fig. 5, table).

Finally, the aggregated data of the previous engineering stages are required for the control engineering. To complete the description, variables, input and output mapping and the list of functions are registered in the sub-elements of the model, e.g. sub elements I/O-mapping and variables of the main element Module_interface, and stored in the AML data structure. Consequently, the completed model information can be used for code generation, e.g. PLCopen XML, and commissioning as well as for documentation.

The benefit of the presented approach is mainly related to the redundant work load in the current kind of engineering process and the information exchanged among the engineering disciplines. Hence, these two impact factors have been measured. Therefore, the percentage of redundant work in current engineering processes has been determined by a survey of experts in the field of material flow systems at the "Materialflusskongress 2016[1]". The results are based on the answers of 23 participants responsible for projects in the field of material flow systems engineering. As a result, the redundancy could be quantified to a range of 10% to 60%. Hence, the consistent description of an aMFM, where each engineer can access the required information and contribute to the discipline-specific information, can significantly increase the engineering efficiency. Along with the redundant information, the dependencies and proportions of the engineering process have been evaluated. Therefore, the information, which can be generated by the appropriate tools as well as the required information for performing an engineering step, have been considered (see Fig. 6). The data is based on the introduced T-junction example that was designed considering the related tools (see Fig. 6). Hence, the strongest dependencies exist between electrical engineering and software engineering (22% of exchanged information) as well as between material flow planning and software engineering (20% of exchanged information). Along with the interfaces, the work proportion of the related disciplines has been measured. A relatively evenly distribution between material flow planning and electrical engineering (about 36% of the process) could be quantified. However, the highest proportion has to be performed by the software engineers (about 37% of the process). These values

---

[1] 25th annual material flow congress 2016, 17th of March 2016, Garching, Germany

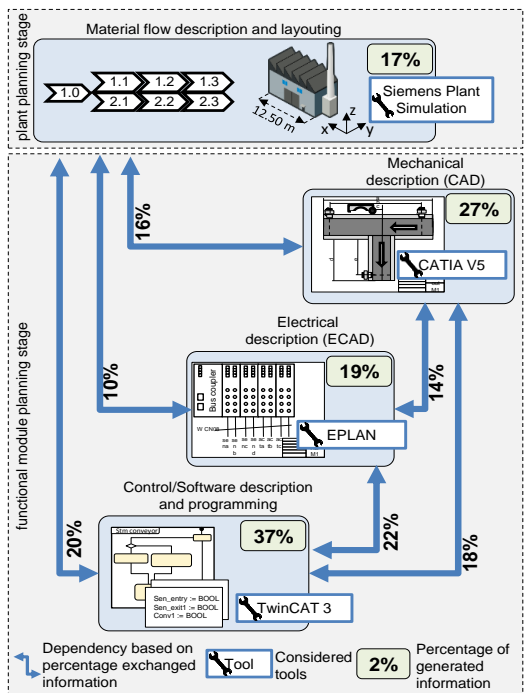

Fig. 6: Evaluation regarding information dependencies during the engineering cycle

could be confirmed by the results of the survey regarding the workload distribution.

Subsequently, the connection of the different stages and disciplines during the engineering cycle, based on the knowledge of the module architecture and a formalized meta model specification, represents the main advantage of our approach. Hence, the required information for the engineer at a certain stage can be provided through the model. Consequently, the effort for an information exchange can be reduced and data consistency increases. However, the tool interfaces have to be adopted to a certain engineering tool chain that entails a considerable effort for implementation. Additionally, a server infrastructure and the prerequisites for accessing the linked documents, e.g. software tools, are required for applying the model-based description. These disadvantages are mainly related to the implementation and, hence, cause only additional effort at one time. However, the advantages regarding decreased redundancy as well as the increased consistency significantly increase the efficiency and quality of an engineering process to at least 10% and up to 60%, which outweighs the effort for implementation.

Additionally, the meta model can only use existing role-classes of the AML. Since the role-class-libraries of the AML are still under development and not completely specified yet, the current scope of application is limited to automatic static material flow modules for piece goods. Other types of material flow systems can be added according to the further definition of the AML standard. However, the upcoming implementation of the AML interface into an increasing number of engineering tools facilitates the integration of our approach, based on the meta model, for the engineering cycle.

Considering the mentioned pros and cons, an initial effort for the implementation of the meta model based approach exists, and the data model relies on the ongoing development of AML. However, the evaluation of the dependencies among the disciplines of the engineering cycle, which were proved by a survey, shows the potential of our approach for increasing the quality and efficiency of the aMFS engineering process.

VI. CONCLUSION AND OUTLOOK

Considering the software engineering in automated production systems, the automation software has to consist of a flexible adaptable structure in the case of extending, reducing or modifying parts of the aMFS. Consequently, the information and documents of an aMFS are often reused during (re-)engineering and need to be completed and structured according to the system architecture in order to reduce the effort and improve error-proneness, e.g. data inconsistency, in the software (re-)engineering. To deal with these requirements and improve the re-use of automation software, modular software architectures are applied. Therefore, the engineering of aMFS integrating different engineering disciplines, e.g. mechanical, electrical/electronic or software engineering, can be broken down into the engineering of small subsystems –modules–. Along with the different engineering disciplines, different engineering stages, e.g. plant planning, functional engineering or commissioning/production, are also addressed by the reference model for the engineering process of aMFS proposed by AML. Considering the AML reference model, there is a high dependency of information and descriptions between these different stages, but an approach to transfer and organize them is lacking. Hence, for instance, information from plant planning to functional engineering is transferred both domain-specific and redundantly, which increases error-proneness and the inconsistency of information.

This paper presented a model-driven design approach for aMFS based on an introduced meta model AutoMFM [31] to improve the transferability of information and description between the engineering stages, plant planning and function engineering. Hence, information from plant planning, which is necessary for the development of functions for aMFM, can be stored consistently in the introduced meta model. Based on this meta model, domains in the plant planning stage can store information necessary for function engineering with the objective of reducing the effort to collect information shared over different domains during the plant engineering process. Additionally, the transferability and consistency of information between different domains in function engineering, e.g. mechanical engineering or control engineering, can be improved by the meta model for aMFM, as shown by an associated survey. Further, based on this meta model, consistency checks of module based information, which improve additionally error-proneness, are enabled.

Future work will examine the measurement of effort, which can be reduced based on the application of a meta model compared to further engineering approaches applied these days. Therefore, special metrics will be applied within industrial case studies. Additionally, further industrial logistic systems will be modelled by the reference model for the engineering process of aMFS proposed by AML to evaluate and, if necessary, extend the introduced meta model AutoMFM.

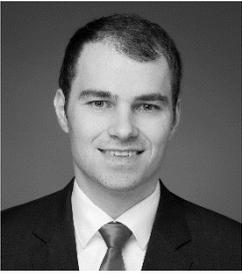

**Daniel Regulin** received his Bachelor degree from the University of Applied Science Hof in 2010 and his Master degree from the Technical University of Clausthal in 2012. Currently, he is a Ph.D. candidate at the Institute of Automation and Information Systems at the Technical University of Munich. His research focus is the model driven engineering of distributed systems, which are organized by multi-agents. Additional research interests are the simulation based analysis mechatronic systems.

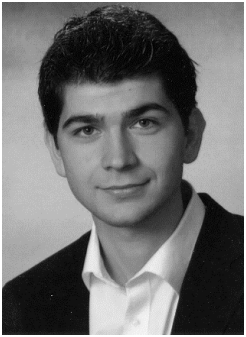

**Thomas Aicher** received the B.Eng. degree in electrical engineering from Deggendorf Institute of Technology (DIT), Deggendorf, Germany, in 2011, and the M.Sc. degree in electrical engineering from Munich University of Applied Sciences (HM), Munich, Germany, in 2013. Currently, he is a Ph.D. candidate at the Institute of Automation and Information Systems (AIS) at the Technical University of Munich (TUM). His research interests include model-driven engineering, verification and design patterns of automation control in the field of automation production systems and material flow systems.

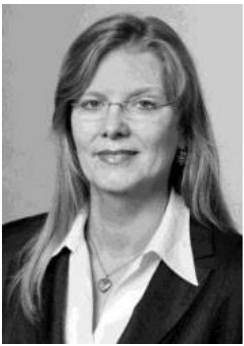

**Birgit Vogel-Heuser** graduated in Electrical Engineering and received the PhD. degree in Mechanical Engineering from the RWTH Aachen in 1991. She worked for nearly ten years in industrial automation in the machine and plant manufacturing industry. After holding different chairs of automation in Hagen, Wuppertal and Kassel she has since 2009 been head of the Automation and Information Systems Institute at the Technical University of Munich. Her main research interests are systems and software engineering, and modeling of distributed and reliable embedded systems. She is coordinator of the Collaborative Research Centre SFB 768: managing cycles in innovation processes – integrated development of product-service systems based on technical products.